\documentclass[12pt]{iopart}

\usepackage{iopams}
\usepackage{graphicx}
\usepackage{bm}
\usepackage{txfonts}
\usepackage{booktabs}

\newcommand{\lco}{LaCoO$_3$}
\newcommand{\leco}{La$_{1-x}$Eu$_x$CoO$_3$}
\newcommand{\rco}{$R$CoO$_3$}
\newcommand{\eco}{EuCoO$_3$}
\newcommand{\pco}{PrCoO$_3$}
\newcommand{\nco}{NdCoO$_3$}
\newcommand{\Cod}{Co$^{3+}$}

\newcommand{\eg}{$e_{g}$}
\newcommand{\DCo}{\ensuremath{\Delta_{\rm Co}}}
\newcommand{\kB}{\ensuremath{k_{\rm B}}}
\newcommand{\Ton}{\ensuremath{T_{\rm on}}}
\newcommand{\as}{\ensuremath{\alpha_{\rm Sch}}}

\begin{document}
\title[Anomalous thermal expansion and damping of the phonon heat transport
 of \boldmath$R$CoO$_{3}$]
{Anomalous expansion and phonon damping due to the Co spin-state
transition in \boldmath$R$CoO$_{3}$ with \bm$R$\,=\,La, Pr, Nd and
Eu}

\author{K~Berggold$^1$, M~Kriener$^1$\footnote{current address:
Department of Physics, Graduate School of Science, Kyoto University,
Kyoto 606-8502, Japan}, P~Becker$^2$, M~Benomar$^1$, M~Reuther$^1$,
C~Zobel$^1$ and T~Lorenz$^1$}

\address{$^1$II. Physikalisches Institut, Universit\"{a}t zu K\"{o}ln,
Z\"{u}lpicher Str.\ 77, 50937 K\"{o}ln, Germany}

\address{$^2$Institut f\"{u}r Kristallographie,
Universit\"{a}t zu K\"{o}ln, Z\"{u}lpicher Str.\ 49b, 50674 K\"{o}ln,
Germany}

\ead{tl@ph2.uni-koeln.de}

\begin{abstract}
We present a combined study of the thermal expansion and the
thermal conductivity of the perovskite series \rco\ with $R=$~La,
Nd, Pr and Eu. The well-known spin-state transition in \lco\ is
strongly affected by the exchange of the $R$ ions due to their
different ionic radii, i.\,e.\ chemical pressure. This can be
monitored in detail by measurements of the thermal expansion,
which is a highly sensitive probe for detecting spin-state
transitions. The Co ions in the higher spin state act as
additional scattering centers for phonons, therefore suppressing
the phonon thermal conductivity. Based on the analysis of the
interplay between spin-state transition and heat transport, we
present a quantitative model of the thermal conductivity for the
entire series. In \pco, an additional scattering effect is active
at low temperatures. This effect arises from the crystal field
splitting of the $4f$ multiplet, which allows for resonant
scattering of phonons between the various $4f$ levels.
\end{abstract}

\pacs{65.40.-b, 65.40.De, 65.40.G-, 72.20.-i}


\maketitle
\section{Introduction}

Cobalt compounds are of particular interest due to the possibility
that Co ions can exhibit different spin states and hence the
occurrence of temperature-driven spin-state transitions. The most
prominent example is \lco, which has been intensively studied and
controversially debated for more than fifty years, see e.\,g.\
\cite{jonker53a,goodenough65a,senaris95a,saitoh97b,asai98a,tokura98a,yamaguchi97a,kobayashi00b,sato08a}.
The Co$^{3+}$ ions in \lco\ feature a $3d^6$ configuration which
in principle can occur in three different spin states: a
nonmagnetic low-spin (LS) ($t_{2g}^{6}e_{g}^{0}$, $S=0$), an
intermediate-spin (IS) ($t_{2g}^{5}e_{g}^{1}$, $S=1$) and a
high-spin (HS) state ($t_{2g}^{4}e_{g}^{2}$, $S=2$). It is
generally agreed that the \Cod\ ions in \lco\ realize the LS state
at low temperatures. Above approximately 25\,K a higher spin
state, either IS or HS, becomes thermally populated affecting
various physical properties, e.\,g.\ the magnetic susceptibility
$\chi$ or the thermal expansion $\alpha$, which both exhibit
pronounced maxima in their temperature dependencies
\cite{zobel02a,baier05a}. The susceptibility is obviously affected
because the excited spin state, either IS or HS, induces a strong
increase of the magnetization above 25\,K. The thermal expansion
is affected due to the different ionic radii of the smaller
LS-\Cod\ with empty and the larger IS- or HS-\Cod\ ions with
partially filled $e_{g}$ orbitals. The spin-state transition can
be well described in a LS\,--\,IS scenario, i.\,e.\ the excited
spin state is the IS state, with a constant energy gap of
$\DCo=185$\,K \cite{zobel02a,baier05a}. However, more recent
investigations show, that a LS/HS model including spin-orbit
coupling is more reasonable
\cite{noguchi02a,haverkort06a,podlesnyak06a}. A consequence of the
latter model is a temperature-dependent energy gap \DCo(T), which
strongly increases with increasing temperature
\cite{haverkort06a}.

The spin-state transition in \lco\ is strongly affected by both,
heterovalent and isovalent doping on the La site. The former
possibility causes hole doping and chemical pressure and
suppresses the spin-state transition due to the implementation of
Co$^{4+}$ ions \cite{kriener04a,kriener08a}. The latter one,
i.\,e.\ chemical pressure without changing the Co valence, is
usually realized by introducing trivalent rare-earth ions
$R^{3+}$. In \rco\ the spin-state transition is not suppressed but
its onset is shifted to higher temperature. The energy gap between
the LS and the excited spin state increases from about
$\DCo=185$\,K for $R=$~La to $\gtrsim 2000$\,K for $R=$~Eu
\cite{baier05a}. Moreover, due to the decreasing ionic radius of
the lanthanide series, the structure changes from rhombohedral in
\lco\ to orthorhombic for $R={\rm Pr}$, Nd and Eu. Recently, it
has been reported, that the low-temperature thermal conductivity
of \lco\ is also very anomalous \cite{yan04a,berggold05a}. This
behaviour has been qualitatively attributed to the onset of the
spin-state transition. However, a quantitative analysis of the
anomalous thermal conductivity of \lco\ has not been presented
yet.

The aim of this paper is to develop a consistent picture of the
influence of the spin-state transition on the thermal
conductivity. Therefore, we measured the thermal conductivity
$\kappa(T)$ of the series \rco\ with $R=$~La, Pr, Nd and Eu.
Moreover, we studied the thermal expansion $\alpha(T)$, which is a
very sensitive probe to investigate spin-state and also
crystal-field transitions and their coupling to the lattice. In
the quantitative analysis of our data, we will consider both
models of the spin-state transition, which are favoured for \lco\
in the literature, i.\,e.\ we will consider the LS\,--\,IS
scenario with a constant \DCo\ and the more recently proposed
spin-orbit coupled HS (SOcHS) model with a temperature dependent
energy gap $\DCo (T)$.

\section{Experiment}
\begin{table}
\caption{Characteristic properties of the investigated crystals.
Here, \Ton\ denotes the onset temperature of the spin-state
transition, \DCo\ the energy gap between the LS ground state and
the excited spin state (either IS or HS) and $\gamma$ is related
to their ionic radii difference (see text). Due to the large \Ton\
of \eco\ we can only give lower limits for the values of $\gamma$
and \DCo, as it was also the case in the related analysis of the
magnetic susceptibility \cite{baier05a}. The last three columns
give the room-temperature values of the thermopower $S_{\rm RT}$,
which for the \lco\ crystals have been taken from
\cite{berggold05a}, and the crystal dimensions (sample cross
section $A$ and sample length $L$) which are important for the
measurements of $\kappa$ (the cross section of S5 is an
approximate value, since it is not of rectangular
shape).\label{tab1}}
\begin{indented}
\item[]\begin{tabular}{lrrrrcr}
\br
Sample & \Ton\ (K) & \DCo\,(K) & $\gamma$ (\%) &$S_{\rm RT}$\,($\mu$V/K) & $A$\,(mm$^2$) & $L$\,(mm) \\
\mr
\lco\ (S1) &  25 &  185 & 0.7   & -700 & $0.8\times1.5$  & 3.7 \\
\lco\ (S2) &     &      &       & 1000 & $1.1\times0.8$  & 2.3 \\
\lco\ (S3) &     &      &       & -600 & $1.4\times0.8$  & 3.4 \\
\lco\ (S4) &     &      &       & 1000 & $0.8\times1.4$  & 2.6 \\
\lco\ (S5) &     &      &       & -300 & $\simeq 2.3$    & 4.7 \\
\pco\      & 175 & 1200 & 2.8   & -400 & $1.0\times3.8$  & 2.5 \\
\nco\      & 230 & 1700 & 4.8   & -400 & $1.7\times2.0$  & 2.8 \\
\eco\      & 400 & $\gtrsim 2000$ & $\gtrsim 10$ & -500& $0.3\times0.85$ & 2.0 \\
\br
\end{tabular}
\end{indented}
\end{table}

All \rco\ crystals have been grown in a floating-zone image
furnace. We examine five different \lco\ crystals identical to
those used in \cite{berggold05a}, where details of the sample
preparation and characterization are given. For \eco\ this
information can be found in \cite{baier05a}. The \nco\ and \pco\
single crystals have been grown in the same way as those of \lco\
and \eco. Characteristic properties of all crystals are listed in
\tref{tab1}. The thermal conductivity  measurements have been
performed by a standard steady-state method using a differential
Chromel-Au+0.07$\,\%\,$Fe-thermocouple \cite{berggold06a}. The
thermal expansion below $\approx 200$\,K was measured using a
home-built high-resolution capacitance dilatometer \cite{pott83a},
whereas the high-temperature measurements $135\,{\rm K} \lesssim T
\lesssim 670$\,K were performed using commercial inductive
dilatometer (TMA 7, Perkin-Elmer).

\section{Results and Discussion}
\subsection{Thermal Expansion}
\begin{figure}
\hfill
\includegraphics[width=0.85\columnwidth,clip]{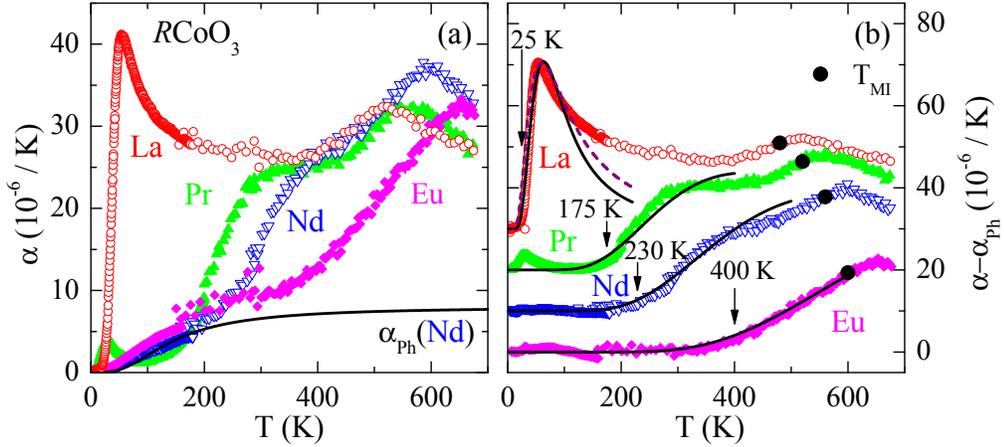}
\caption{(a) Thermal expansion of \rco\ with $R=$~La, Pr, Nd and
Eu. The solid line is the estimated phonon contribution of \nco\
(see text). (b) Anomalous part of the thermal expansion obtained
by subtracting the phononic background. For clarity the different
data sets are offset by $10^{-5}$\,/\,K with respect to each
other. The arrows signal the approximate onset temperature of the
spin-state transition and the solid circles denote the
metal-insulator transition temperature estimated from resistivity
measurements for each compound. The solid lines are fits of the
respective Schottky anomalies, see equations\,(\ref{schott})
and~(\ref{schott2}), assuming a constant energy gap \DCo\ between
the different spin states of \Cod. The dashed line for \lco\ is a
similar fit using $\DCo (T)$ obtained from the magnetic
susceptibility (see text).} \label{TADRCoO3}
\end{figure}

In \fref{TADRCoO3}\,(a) we show the thermal-expansion coefficients
$\alpha(T)=1/L\cdot\partial L/\partial T$ of \rco. The
low-temperature results of \lco\ and \eco\ were already discussed
in detail in \cite{zobel02a,baier05a}. The thermal expansion
consists of a phononic part and a Schottky contribution, caused by
the thermal population of the $e_g$ orbitals of the Co$^{3+}$
ions, i.\,e.\ $\alpha=\alpha_{\rm Ph}+\as$. The latter
contribution causes the large maximum at $\simeq 50$\,K in \lco.
To further analyze the data we subtract $\alpha_{\rm Ph}$, which
we estimate using a Debye function with the Debye temperature
$\Theta_{\rm D}=600$\,K of \lco\ \cite{stolen97a}. Here, we used
the same $\Theta_{\rm D}$ for the various cobaltates, but
sample-dependent prefactors determined by scaling the Debye
function to the low-temperature data of each compound. As an
example, we show $\alpha_{\rm Ph}$ for \nco\ in
\fref{TADRCoO3}\,(a); $\alpha_{\rm Ph}$ of \eco\ (\pco; not shown)
is slightly larger (smaller). Since a clear separation of
$\alpha_{\rm Ph}$ and $\as$ is not possible for the
low-temperature data of \lco, we used $\alpha_{\rm Ph}$ of \eco\
also for \lco. The resulting $\as$ of all crystals are shown in
\fref{TADRCoO3}\,(b). We note that, in particular for \pco\ and
\nco\ the spin-state transition is seen much better in the
thermal-expansion data than in the magnetic susceptibility. The
reason is that $\alpha_{\rm Ph}$ is rather small compared to the
total thermal expansion, whereas $\chi$ is dominated by the large
contribution of the $4f$ moments of the Pr$^{3+}$ and Nd$^{3+}$
ions, respectively, which makes a further analysis rather
uncertain. The insulator-metal transitions occurring above about
500\,K, see e.\,g.\ \cite{baier05a,yamaguchi96b} also cause
anomalies in $\alpha(T)$. The solid circles in
\fref{TADRCoO3}\,(b) signal the transition temperatures $T_{\rm
MI}$, which have been determined from resistivity measurements on
our crystals (not shown).

Within the LS/IS scenario with a constant \DCo, $\as(T)$ is given
by \cite{zobel02a}
\begin{equation}
\as(T)=\gamma\;\frac{\DCo}{T^2}\;\frac{\nu\exp(-\DCo/T)}
{\left(1+\nu\exp(-\DCo/T)\right)^2}\, , \label{schott}
\end{equation}
where $\nu=3$ is the degeneracy of the excited spin state and
$\gamma $ is a measure of the ionic radii difference of the \Cod\
in the LS and the excited spin state. Moreover, $\gamma$ is
related to the uniaxial pressure dependence of $\DCo$ via
$\gamma=\kB/V_{\rm fu}\cdot \partial \DCo/\partial p_\alpha$,
where \kB\ is Boltzmann's constant, $V_{\rm fu}\simeq 56$\,\AA$^3$
the volume per formula unit and $p_\alpha$ means uniaxial pressure
along the direction of which $\alpha$ is measured
\cite{lorenz07a,lorenz08a}. To get the hydrostatic pressure
dependence one has to use the volume expansion which in the case
of the twinned \rco\ single crystals is given by $3\,\alpha$. The
corresponding fit of $\alpha$ of \lco\ (already presented in
\cite{zobel02a}) is shown by the solid line in
\fref{TADRCoO3}\,(b). The fit parameters are $\DCo=185$\,K and
$\gamma = 0.007$ giving a hydrostatic pressure dependence
$\partial (\ln \DCo) / \partial p_{\rm hydr}\simeq 45\,\%/\,{\rm
GPa}$, in agreement with the increase of \DCo\ of $\simeq 42$\,\%
obtained from measurements of $\chi$ under $\simeq 1.1$\,GPa
\cite{asai97a}.

For the SOcHS model one has to consider the temperature dependence
of \DCo\ \cite{haverkort06a}, which can be calculated from the
measured $\chi(T)$ \cite{baier05a}. For \lco, this yields a linear
increase $\DCo^{\rm HS}(T)=\DCo^{0}+a\cdot T$ with
$\DCo^{0}=135$\,K and $a=1.66$ in the temperature range almost up
to the MI transition. Using this $\DCo^{\rm HS}(T)$ in the
partition sum, equation\,(\ref{schott}) modifies to
\begin{equation}
\as(T)=\gamma^{\rm HS}\;
\frac{\DCo^0}{T^2}\frac{\nu\exp(-(\DCo^0+a\,T)/T)}
{\left(1+\nu\exp(-(\DCo^0+a\,T)/T)\right)^2}\:.
\label{schott2}
\end{equation}
The corresponding fit of $\as$ with $\gamma^{\rm HS}$ as the only
free parameter gives the dashed line in \fref{TADRCoO3}\,(b).
Obviously, both fits hardly differ because the modified $\DCo^{\rm
HS}$ is compensated by a larger value of $\gamma^{\rm HS}=0.02$.
We note, however, that this does {\em not} necessarily correspond
to a larger pressure dependence of $\DCo^{0}$, because $\as$ of
equation\,(\ref{schott2}) is determined by the pressure
dependencies of $\DCo^{0}$ and that of the slope $a$; see e.\,g.\
the discussions in \cite{lorenz07a,johannsen05a,anfuso08a}. In
view of the small differences between both fits for \lco, and
since there are no indications for a temperature-dependent \DCo\
from other physical quantities, the fits of $\as$ of the other
\rco\ crystals have been done for constant energy gaps only. Since
equations\,(\ref{schott}) and~(\ref{schott2}) are identical for
$a=0$, this analysis is not able to distinguish between both
models. The obtained values of \DCo\ and $\gamma$ are given in
\tref{tab1}. As expected, \DCo\ strongly increases with decreasing
radius of the $R^{3+}$ ions and therefore the spin-state
transition monotonically shifts to higher temperature. The onset
temperature \Ton\ of the spin-state transition for each compound,
$\Ton\simeq 25$, 175, 230 and 400\,K for $R=$~La, Pr, Nd and Eu,
respectively, are marked by arrows in \fref{TADRCoO3}\,(b).

\subsection{Thermal conductivity}
\begin{figure}
\hfill
\includegraphics[width=0.85\columnwidth,clip]{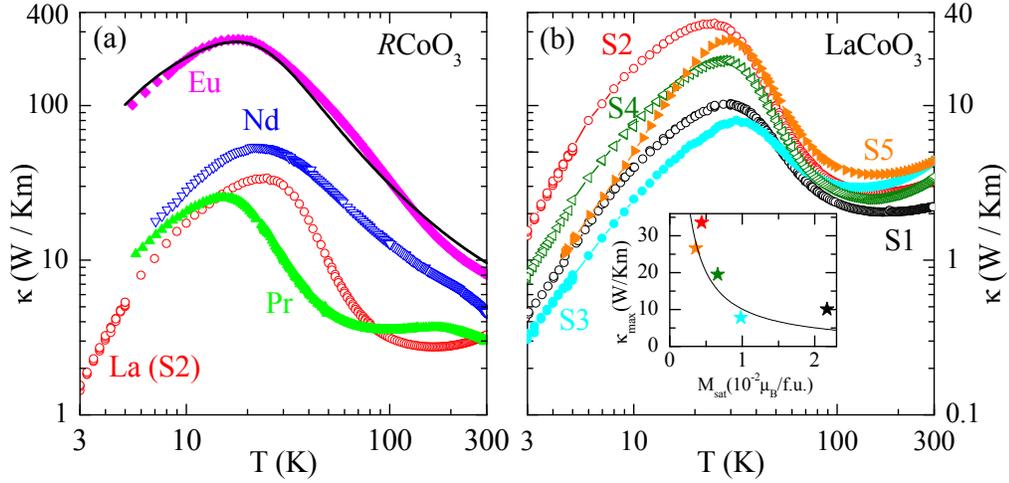}
\caption{(a) Thermal conductivity of \rco\ with $R\,=$\,La, Pr, Nd
and Eu; the solid line is a fit via equation\,(\ref{kphonon}). (b)
Thermal conductivity of five different \lco\ single crystals
S1\,--\,S5 (note the different scale). The inset shows the maximum
values of the thermal conductivity $\kappa_{\rm max}$ of these
crystals vs. the saturation values of the magnetization $M_{\rm
sat}$ measured at 2\,K. As shown by the sold line, we find an
approximate proportionality $\kappa_{\rm max}\propto 1/M_{\rm
sat}$ (see text).} \label{kappaRCoO3}
\end{figure}

\Fref{kappaRCoO3}\,(a) displays the thermal-conductivity data of
\rco. As has been already found in previous studies on \lco\
\cite{yan04a,berggold05a}, the overall shape and qualitative
behaviour of $\kappa(T)$ is rather unusual. First of all, $\kappa$
is rather low in the whole temperature range and its temperature
dependence clearly deviates from the typical behaviour expected
for a conventional phononic heat conductor. The thermal
conductivity of \lco\ exhibits a maximum around 20\,--\,30\,K.
Towards higher temperatures $\kappa$ rapidly drops and features a
minimum around 150\,K instead of the expected $1/T$-decrease (see
also \cite{berggold05a}). Qualitatively, this minimum can be
traced back to the spin-state transition of \lco, which sets in
close to the maximum of $\kappa$. The thermal population of the Co
$e_g$ levels induces a certain fraction of Co$^{3+}$ ions with a
larger ionic radius compared to the LS \Cod\ ions. These randomly
distributed larger \Cod\ ions do not only cause the huge anomaly
in $\alpha(T)$, but also lead to additional disorder in the
lattice and therefore act as additional scattering centers for
phonons. Hence, the thermal conductivity is additionally
suppressed. A quantitative description will be given below.

Since the spin-state transition shifts to higher temperature when
La$^{3+}$ is replaced by smaller $R^{3+}$ ions, the above scenario
suggests that the onset of the suppression of $\kappa(T)$ of \rco\
systematically shifts towards higher temperature with decreasing
radius of the $R^{3+}$ ion. Let us first consider the data of
\eco, which is the compound with the smallest $R^{3+}$ ion and the
largest $\DCo\gtrsim 2000$\,K \cite{baier05a}. Hence, the
spin-state transition should not affect the thermal conductivity
below room temperature and, indeed, the thermal conductivity of
\eco\ exhibits much larger values than in \lco. Its
low-temperature maximum exceeds that of \lco\ by about one order
of magnitude. Moreover, the temperature dependence of $\kappa$ of
\eco\ meets the expectation of a purely phononic heat conductor.

The thermal conductivity of \nco\ is lower than $\kappa$ of \eco\
in the entire temperature range and, at a first glance, its
overall temperature dependence seems to be consistent with a
conventional phononic picture, too. A closer inspection of the
data yields, however, a slight slope change towards a somewhat
steeper decrease of $\kappa(T)$ at $\approx 230$\,K, i.\,e.\ close
to the onset temperature of the spin-state transition, see
\fref{TADRCoO3}\,(b). This slope change supports the above
proposed explanation that the spin-state transition causes
additional scattering centers.

\pco\ also exhibits an unusual temperature dependence of $\kappa$,
which appears somewhat similar to that of \lco. Above the
low-temperature maximum, $\kappa(T)$ rapidly decreases to a small
value of $\simeq 3.5$\,W/Km and remains almost temperature
independent between about 80\,K and 175\,K. Above this temperature
a further decrease of $\kappa$ sets in. Since this decrease starts
again at the onset of the spin-state transition (see
\fref{TADRCoO3}\,(b)), it is natural to attribute it to the
additional scattering due to the presence of \Cod\ ions in
different spin states. In contrast, the anomalous temperature
dependence of $\kappa(T)$ at lower temperature is not related to
the spin-state transition and, as will be discussed separately in
\sref{sec_prcoo3res}, this anomalous feature is related to the
crystal-field splitting of the $4f$ shell of Pr$^{3+}$.

Summarizing the presented data so far, we observe in all crystals
with $T_{\rm on}<300$\,K that this onset coincides with an
additional suppression of $\kappa$. Since \DCo\ strongly increases
with decreasing $R^{3+}$ radius \cite{baier05a,yamaguchi96b},
\Ton\ also shifts towards higher temperature and the additional
suppression of the thermal conductivity is less obvious, because
at higher temperature the usual suppression of $\kappa$ due to
phonon-phonon Umklapp scattering becomes more and more dominant.
However, the data in \fref{kappaRCoO3}\,(a) also show, that
$\kappa$ is not only suppressed at temperatures {\it above} \Ton.
Instead, with increasing size of $R^{3+}$ a monotonic decrease of
$\kappa$ in the \textit{entire} temperature range is found (here
we have neglected the additional suppression of $\kappa $ in
\pco). If the spin-state transition would be the only effect
causing additional disorder scattering, $\kappa(T)$ should
approach the 'bare' phononic value below \Ton. Taking \eco\ with
stable LS Co$^{3+}$ ions up to above room temperature as a rough
measure for this 'bare' phononic heat conduction, it becomes clear
that this expectation is not observed in our data.

In general, the magnitude of the low-temperature maximum of
$\kappa$ depends on the sample quality, which determines the
maximum free path. Thus, the different values within the \rco\
series could arise from different sample qualities. However, there
is no indication for a systematic increase of the crystal quality
with decreasing $R^{3+}$ ion radius.\footnote{One may speculate
that the low-temperature maximum of $\kappa$ is also influenced by
twinning. All these single crystals are twinned, but it seems
plausible that the average size of the twin domains increases from
LaCoO$_{3}$ to EuCoO$_{3}$, because the deviation from the cubic
symmetry increases. However, the observed correlation between
$\kappa_{\rm max}$ and the magnetic impurity concentration for the
different LaCoO$_{3}$ crystals, makes it rather unlikely to
explain the different low-temperature maxima of $\kappa$ in terms
of different domain sizes.} In order to study a possible influence
of the crystal quality, we measured $\kappa$ of five different
\lco\ single crystals. As shown in \fref{kappaRCoO3}\,(b) the
high-temperature behaviour is essentially the same for all
crystals, but the low-temperature maxima differ considerably.
Nevertheless, all of them remain much lower than that of \eco. We
relate the low-temperature maxima of $\kappa$ to the defect
concentrations by considering the low-temperature magnetization,
since a really pure \lco\ with all Co$^{3+}$ ions in the LS state
would be non-magnetic (apart from core dia- and van~Vleck
paramagnetism). At $\simeq 2$\,K we measured magnetization up to
14~Tesla in order to reach its saturation $M_{\rm sat}$, which is
a measure of the content of magnetic impurities. The inset of
\fref{kappaRCoO3} displays the low-temperature maxima of
$\kappa(T)$ vs. $M_{\rm sat}$, which indeed roughly follow a
$1/M_{\rm sat}$ dependence as shown by the solid line.

The most likely source for magnetic impurities in \lco\ is a weak
oxygen nonstoichiometry, which changes the average valence from
Co$^{3+}$ towards Co$^{2+}$ or Co$^{4+}$. Due to the odd number of
$3d$ electrons, the latter are magnetic in all possible electronic
configurations, but in most cases Co$^{2+}$ and Co$^{4+}$ realize
$t_{2g}^5e_g^2$ and $t_{2g}^5e_g^0$ states with spin $S=3/2$ and
$S=1/2$, respectively. Experimental evidence for a weak
charge-carrier doping in nominally pure \lco\ stems from the large
values of different signs of the room-temperature thermopower for
the different \lco\ crystals; cf.\ \tref{tab1} and the discussion
in \cite{berggold05a}. From magnetization data on Sr-doped \lco\
it is known that low doping concentrations induce so-called
high-spin polarons with effective spin values up to $S=16$
\cite{yamaguchi96a}. Such polarons are formed around the
doping-induced Co$^{4+}$ ions, which cause a spin-state transition
in the adjacent Co$^{3+}$ ions. Most probably, this is a
transition from the LS to an IS state, because the electronic
configuration $t_{2g}^5e_g^1$ of Co$^{3+}$ is favourable for the
hopping of the $e_g$ electron for both kinds of neighbours,
Co$^{4+}$ with $t_{2g}^5e_g^0$ and Co$^{2+}$ with $t_{2g}^5e_g^2$.
Other possible configurations often only allow a $t_{2g}$ hopping
or may cause the so-called spin-blockade effect
\cite{maignan04b,lengsdorf04a}. The formation of high-spin
polarons will depend on \DCo\ between the LS ground state and the
excited spin state of \Cod. Since \DCo\ strongly increases with
decreasing $R^{3+}$ radius, the occurrence of such polarons in the
\rco\ series should get more and more unlikely for smaller
$R^{3+}$ ions. Unfortunately, the presence of high-spin polarons
in \pco\ and \nco\ cannot be studied by magnetization measurements
because of the strong magnetism of the partially filled $4f$
shells of Pr$^{3+}$ and Nd$^{3+}$. For \eco\ such an analysis is
again easier, because Eu$^{3+}$ only exhibits van~Vleck
paramagnetism and, in fact, the impurity contribution to the
magnetic susceptibility in \eco\ is one order of magnitude smaller
than in \lco\ \cite{baier05a}. Moreover, the magnetic impurity
contribution of the \leco\ series systematically decreases with
increasing $x$, i.\,e.\ with decreasing average ionic size on the
La site and with increasing \DCo\ \cite{baier05a}. These
observations strongly support the above idea that the systematic
dependence of the low-temperature thermal conductivity of \rco\ on
the ionic size of $R^{3+}$ is related to phonon scattering by
high-spin polarons, whose formation depend on a weak
oxygen-nonstoichiometry and on \DCo.

\section{Quantitative analysis of the thermal conductivity}
\label{quantis} For a quantitative analysis of the thermal
conductivity we use an extended Debye model \cite{bermann76a}
\begin{equation}
\kappa(T) = \frac{\kB^4 T^3}{2 {\pi}^2 \hbar^3 v_{\rm s}}
\int\limits_{0}^{\Theta_{\rm D}/T} \tau(x,T) \frac{x^4 e^x} {( \rme^x - 1)^2}{\rm d}x. \label{kphonon}
\end{equation}
Here, $\Theta_{\rm D}$ denotes the Debye temperature, $v_{\rm s}$ the sound velocity, $\omega$ the phonon frequency, $x = \hbar\omega/k_{\rm B}T$ and $\tau(x,T)$ the phonon relaxation time. The scattering rates of different independent scattering mechanisms sum up to a total scattering rate
\begin{equation}
\tau^{-1}(x,T) = \tau_{\rm bd}^{-1}+\tau_{\rm pt}^{-1} +\tau_{\rm um}^{-1} = \frac{v_{\rm s}}{L} + P \omega^4 + U T \omega^3 \exp(\frac{\Theta_{\rm D}}{uT}).
\label{rates}
\end{equation}
The three terms on the right-hand side refer to the typical
scattering rates for phonon heat transport in insulators, namely
boundary ($\tau_{\rm bd}$; $L=1$\,mm is the characteristic length
scale of the sample \cite{li07a} and was kept fixed for all
crystals), point defect ($\tau_{\rm pt}$) and phonon-phonon
Umklapp scattering ($\tau_{\rm um}$). Since the lattice spacing is
a natural lower limit for the mean-free path, $\kappa(T)$ is
expected to approach a minimum value at higher temperatures. In
order to model this, $\tau(x,T)$ in equation\,(\ref{rates}) is
replaced by $\max\{\tau_\Sigma(x,T),\ell_{\rm min}/v_{s}$\} with
the minimum mean-free path $\ell_{\rm min}$ being of the order of
the lattice constant; compare the discussion in
\cite{kordonis06a}.

As shown by the solid line in \fref{kappaRCoO3}, the thermal
conductivity data of \eco\ can be reproduced reasonably well by
equation\,(\ref{kphonon}). In order to reduce the number of
adjustable parameters, we used $\Theta_{\rm D}=600$\,K and $v_{\rm
s}=3900$\,m/s determined for \lco\ \cite{stolen97a,murata99a}. The
other parameters have been adjusted to fit the experimental data.
We obtain $P=3.7\cdot10^{-43}$\,s$^3$,
$U=4.9\cdot10^{-31}$\,s$^2$/K and $u=7.6$; $\ell_{\rm min}$ is not
reached in \eco\ up to 300\,K.

\subsection{Scattering related to the spin-state transition}
In the following analysis we consider the thermal conductivity of
\eco\ as the bare phononic heat conductivity that is not
influenced by the spin-state transition. For the other members of
the \rco\ series we introduce an additional scattering mechanism
arising from the thermal occupation of the Co$^{3+}$ \eg\ shells,
which causes additional lattice disorder due to the presence of
smaller and larger Co$^{3+}$ ions in the LS and the excited spin
state, respectively. Considering the \Cod\ ions of different sizes
as point-like defects, we assume an $\omega^4$ frequency
dependence for this additional scattering rate (cf.\
equation\,(\ref{rates}))
\begin{equation}
\tau_{\rm dis}^{-1} = C\cdot f(T)\cdot\omega^4, \label{tauexp}
\end{equation}
where $C$ is an adjustable parameter describing the scattering
strength and $f(T)$ models the temperature dependence of the
additional lattice disorder.

\begin{figure}
\hfill
\includegraphics[width=0.85\columnwidth,clip]{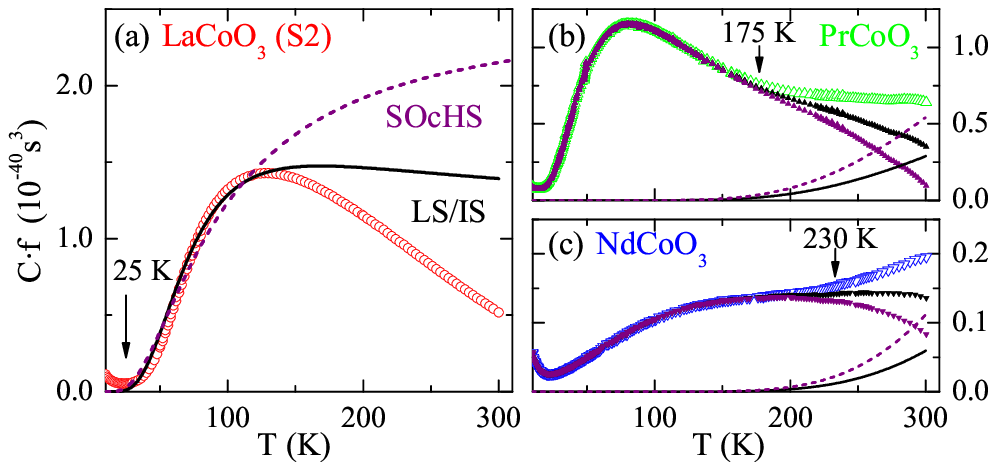}
\caption{Open symbols denote the experimentally obtained
additional scattering rates $C\cdot f(T)$, which are needed to
suppress $\kappa$ from a hypothetical bare phonon heat
conductivity represented by $\kappa$ of \eco\ to the measured
values of $\kappa$ of \lco, \pco\ and \nco (see text for details).
In all three \rco\ an additional scattering sets in above the
respective onset temperature \Ton\ (marked by the arrow in each
panel) of the spin-state transition. The lines show the expected
additional scattering rates due to the spin-state transition for a
LS/IS scenario (solid) and for the SOcHS model (dashed); see
equations~(\ref{tauexp}) and~(\ref{equ_taudis}). In panel (a) the
scattering strengths $C$ for each model have been adjusted by
fitting the lines to the experimental data in the temperature
range below 150\,K. These values have been kept fixed to calculate
the lines in panels~(b) and~(c), and, as shown by the solid
symbols, the additional scattering above \Ton\ completely
vanishes, when these lines are subtracted from the experimental
$C\cdot f(T)$.} \label{tauRCoO3}
\end{figure}

The unknown functions $C\cdot f(T)$ for \lco, \pco\ and \nco\ can
be calculated from the experimental data in the following way: we
use equation\,(\ref{rates}) with all the parameters obtained from
the fit of $\kappa$ of \eco\ kept fix and add $\tau_{\rm
dis}^{-1}$. Then the total $\tau$ is used in
equation\,(\ref{kphonon}) and the values of $C\cdot f(T)$ at each
temperature are determined point by point from the measured
$\kappa(T)$. The resulting $C\cdot f(T)$ curves for all three
\rco\ samples are shown by the open symbols in \fref{tauRCoO3}. In
order to avoid additional complications arising from the
aforementioned high-spin polarons, we restrict the analysis to the
temperature range above $\simeq 25$\,K. For \lco\ the resulting
$C\cdot f(T)$ has a pronounced maximum around 100\,K, and strongly
resembles the temperature dependence of the magnetic
susceptibility of \lco. Here, we have used Sample S2 which
exhibits the highest value of the low-temperature maximum of
$\kappa$, i.\,e.\ the sample which is least affected by impurity
scattering. We note, however, that in the considered temperature
range above 25\,K the resulting $C\cdot f(T)$ is not very
different for the other \lco\ crystals. For \pco , $C\cdot f(T)$
features a clear maximum followed by a slope change around 175\,K.
The obtained values of $C\cdot f(T)$ for \nco\ are much lower than
those of \lco\ and \pco\ and there is no maximum, but again a
clear slope change around $T_{\rm on}\simeq 230$\,K. Thus, we find
in all three \rco\ compounds either an increase of $C\cdot f(T)$
or a slope change towards larger values at temperatures close to
the respective onset temperatures of the spin-state transition;
see \fref{TADRCoO3}.

In order to model the additional scattering rate we use the
so-called Nordheim rule \cite{nordheim31a}, which is a successful
approach to describe $\kappa$ in disordered mixed alloys. There,
the additional scattering rate is proportional to $x(1-x)$, with
$x$ being the fraction of one of the constituents. In the present
case this corresponds to the fractions $n_{\rm IS/HS}$ of \Cod\
ions in the higher and $n_{\rm LS}=(1-n_{\rm IS/HS}$) in the LS
state. Thus, the temperature dependence of the additional
scattering rate of equation\,(\ref{tauexp}) is expected to be
given by (see equations~(\ref{schott}) and~(\ref{schott2}))
\begin{equation}
f(T)=\frac{\nu\exp(-\DCo(T)/T)}{\left(1+\nu\exp(-\DCo(T)/T)\right)^2}\:.
\label{equ_taudis}
\end{equation}

The solid line in \fref{tauRCoO3}\,(a) corresponds to a
calculation of $C\cdot f(T)$ via equation~(\ref{equ_taudis}) using
the temperature-independent $\DCo=185$\,K of the LS\,--\,IS
scenario. Here, $C$ is the only adjustable parameter, and for
$C=5.9\cdot 10^{-40}$\,s$^3$ this model yields a good description
of the experimental data in the temperature range from 30\,K to
150\,K. The dashed line shows the corresponding fit (yielding
$C=11\cdot 10^{-40}$\,s$^3$) for the same temperature range based
on the SOcHS scenario with a temperature-dependent $\DCo (T)$,
which is in reasonable agreement with the data, too. Obviously,
both models are not able to describe the decrease of $f(T)$ above
150\,K. This disagreement does, however, not contradict the
models, because for $T\gtrsim 150$\,K the measured $\kappa$ of
\lco\ is already close to its minimum value arising from the
aforementioned lower limit of the mean-free path. For $C\cdot
f(T)$ of \pco\ and \nco, we keep the values of $C$ fixed and use
the gap values $\DCo=1200$\,K and $\DCo=1700$\,K obtained from the
analysis of $\alpha(T)$, cf.~\tref{tab1}. The corresponding
curves, which for the different models only differ in the
scattering strengths, are shown as solid and dashed lines in
Figs.\,\ref{tauRCoO3}\,(b) and~(c). When we subtract these lines
from the experimental data, the slope changes around \Ton\
completely vanish, as is shown by the solid symbols in
Figs.\,\ref{tauRCoO3}\,(b) and~(c). This gives further evidence
that the spin-state transition is indeed responsible for the
suppression of the thermal conductivity of \pco\ and \nco\ as it
is the case for \lco.

\begin{figure}
\hfill
\includegraphics[width=0.85\columnwidth,clip]{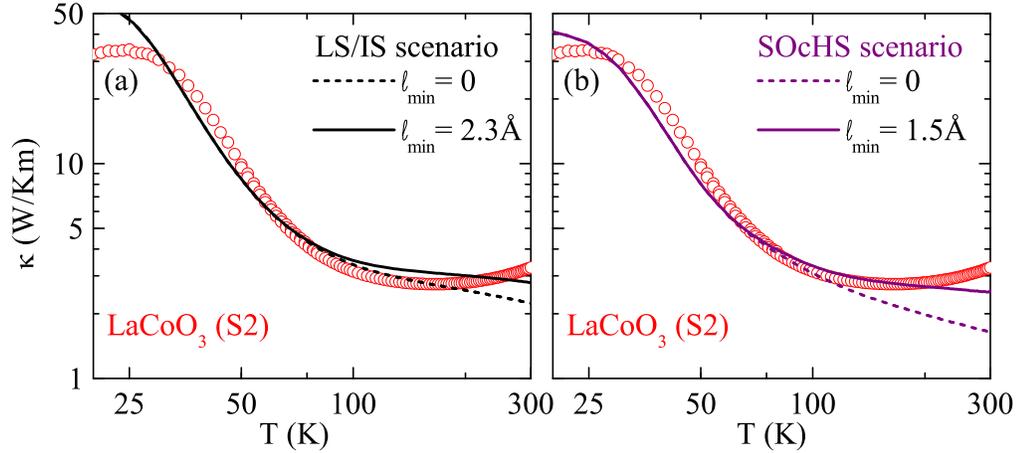}
\caption{Thermal conductivity (symbols) of \lco\ with fits based
on equation\,(\ref{kphonon}) assuming (a) the LS\,--\,IS or (b)
the SOcHS model for the spin-state transition. The dashed (solid)
lines are obtained when a minimum mean-free $\ell_{\rm min}$ is
(is not) taken into account (see text).} \label{kappafitLaCoO3}
\end{figure}

Encouraged by this finding we tried to fit the thermal
conductivity of \lco\ via equation\,(\ref{kphonon}) including the
additional scattering rate given by equations\,(\ref{tauexp})
and~(\ref{equ_taudis}) in equation\,(\ref{rates}).  Again, we
restrict our analysis to $T\gtrsim 25$\,K and keep most of the fit
parameters fixed to their values as obtained from the Debye fit of
the bare phononic $\kappa$ of \eco. From these parameters, we only
readjust the prefactor $P$ of the point-defect scattering term in
equation\,(\ref{rates}), because the number of point defects
varies in different crystals, and point-defect scattering also
reduces $\kappa$ at high temperatures. In addition, the scattering
strength $C$ is readjusted, but changes only little (to $C=6\cdot
10^{-40}$\,s$^3$ and $12\cdot 10^{-40}$\,s$^3$ for the LS/IS and
the SOcHS model, respectively), and we consider the minimum
mean-free path $\ell_{\rm min}$. The results for the LS\,--\,IS as
well as the SOcHS model are displayed in \fref{kappafitLaCoO3}.
Both models describe the thermal conductivity well over a large
temperature range of about 200\,K, in particular, when the minimum
mean-free path is taken into account. As was already seen in
\fref{tauRCoO3} the quality of the fits based on the different
models hardly differs. Therefore, the analysis of the thermal
conductivity of \rco\ does not allow to judge, whether the LS/IS
or the SOcHS scenario is preferable.

\subsection{Scattering related to crystal-field excitations}
\label{sec_prcoo3res} 
\begin{figure}
\hfill
\includegraphics[width=0.85\columnwidth,clip]{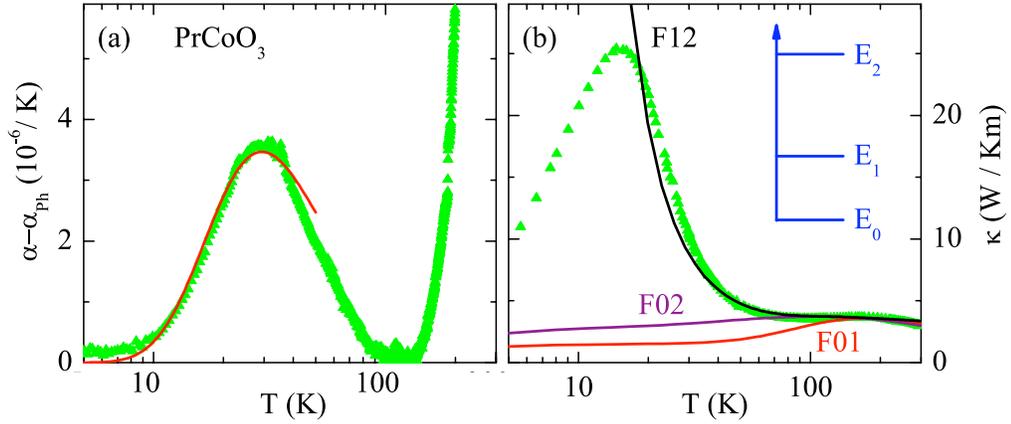}
\caption{Low-temperature (a) thermal expansion and (b) thermal
conductivity of \pco. The thermal-expansion data (symbols) in
panel (a) are well described by a Schottky anomaly (line) of a
two-level system with non-degenerate levels with energy splitting
$E_1-E_0=70$\,K. The lines F01, F02 and F12 in panel (b) are fits
for $T>20$\,K to the thermal conductivity data (symbols) including
either the resonant scattering process $E_0\leftrightarrow E_1$,
$E_0\leftrightarrow E_2$ or $E_1\leftrightarrow E_2$,
respectively. A sketch of the energy levels is also included
\cite{rosenkranz99a}.} \label{lowTPrCoO3}
\end{figure}

Next we consider the additional suppression
of $\kappa$ of \pco\ below \Ton, which is not related to the
spin-state transition. As can be seen in \fref{TADRCoO3}, the
thermal expansion of this compound also exhibits an additional
anomaly at lower temperatures. This anomaly arises from the
crystal-field splitting of the $4f$ shell of the Pr$^{3+}$ ions.
In an orthorhombic crystal field the $^3$H$_4$ multiplet of the
free Pr$^{3+}$ ion splits into nine singlets. We are not aware of
experimental investigations of the crystal-field splitting in
\pco, but it is known that the crystal-field levels of other
orthorhombic Pr$A$O$_3$ are very similar
\cite{feldmann75a,podlesnyak94a,rosenkranz99a}. Only the lower
levels slightly depend on the exact orthorhombic distortion, which
is mainly characterized by the $A$\,--\,O\,--\,$A$ canting angle.
The energy splitting between the crystal-field ground state and
the first excited level in \pco\ can be determined from a fit of
the low-temperature thermal expansion by a Schottky anomaly. As
shown in \fref{lowTPrCoO3}\,(a), such a fit for two non-degenerate
levels (setting $\nu=1$ and replacing \DCo\ by $\Delta_{\rm CF}$
in equation~(\ref{schott})) describes the experimental data well
and yields $\Delta_{\rm CF}=E_1=70$\,K and $\gamma=0.056$~\%. For
the higher-lying energy levels we use the values $E_2=151$\,K,
$E_3=174$\,K, $E_3=235{\rm \,K}, \dots $ of PrNiO$_3$
\cite{rosenkranz99a}, which has almost the same
$A$\,--\,O\,--\,$A$ canting angle as \pco. We note here that the
thermal population of these higher-lying levels may contribute to
$\alpha(T)$ at higher temperatures, but below about 50\,K the
two-level system used is sufficient to model the experimental
data.

The most likely source for the suppression of the thermal
conductivity of \pco\ in the intermediate temperature range is
resonant scattering between different $4f$ crystal-field levels,
as it is also known from e.\,g.\ rare-earth garnets
\cite{slack71a}. In such a resonant scattering process a phonon is
absorbed by inducing a transition between different energy levels
and then re-emitted in an arbitrary direction (see e.\,g.\
\cite{hofmann01a,berggold07a}), causing an additional thermal
resistance. A mechanism based on random volume changes as
discussed above for the spin-state transition of \Cod\ is unlikely
because the $4f$ orbitals are inner shells. This difference is
directly reflected in the magnitudes of the Schottky peaks, which
are mainly determined by the underlying volume change. The
low-temperature peak in $\alpha$ of \pco\ is e.\,g.\ one order of
magnitude smaller than that of \lco; see \fref{TADRCoO3}
and.\footnote{We also mention that our attempts to describe the
additional damping of $\kappa$ due to the spin-state transition of
\lco\ by a resonant scattering process completely failed.}

In order to model $\kappa(T)$ in the low-$T$ range we include  the
rate
\begin{equation}
\tau_{\rm res}^{-1}=R\frac{4\omega^4\Delta_{\rm CF}^4}
{\left(\Delta_{\rm
CF}^2-\omega^2\right)^2}\cdot\left(N_i+N_j\right) \label{equ_res}
\end{equation}
in equation\,(\ref{rates}) describing a direct resonant scattering
process \cite{hofmann01a,orbach61a}. Here, $\Delta_{\rm
CF}=|E_i-E_j|$ is the energy difference between the two levels,
and $N_i$, $N_j$ denote their population determined by the
partition sum. For the fit we keep again $\Theta_{\rm D}=600$\,K
and $v_{\rm s}=3900$\,m/s, while $P$, $U$, $u$ and $R$ are
adjusted with respect to the data. The fits have been restricted
to $T\geq 20$\,K, since other processes may become dominant at
very low temperatures.

In a first attempt, resonant scattering between $E_0$ and the
first excited level $E_1$ is considered, yielding the fit labelled
F01 in \fref{lowTPrCoO3}\,(b). This fit only reproduces the data
for $T\gtrsim 100$\,K but totally misses the low-temperature
maximum in $\kappa(T)$. Basically the same result is achieved by
considering the process $E_0\leftrightarrow E_2$ (curve F02 in
\fref{lowTPrCoO3}\,(b)). A much better description of the data is
obtained, however, for the process $E_1\leftrightarrow E_2$ (curve
F12; with parameters $P=5\cdot 10^{-43}$\,s$^3$, $U=4.85\cdot
10^{-18}$s$^2$/K, $u=5$ and $R=3.8\cdot 10^{-40}$\,s$^3$). Hence,
the transition $E_1\leftrightarrow E_2$ appears to be the dominant
scattering channel of the phonons in \pco\ at intermediate
temperatures.

\section{Summary}
In this paper we have presented measurements of the
thermal-expansion coefficient and the thermal conductivity of
\rco, with trivalent $R=$~La, Pr, Nd and Eu. As a consequence of
the chemical pressure due to the decreasing ionic radius from
$R=$~La to Eu, the energy gap between the LS ground state and the
first excited spin state of the \Cod\ ions increases drastically
and the temperature-induced spin-state transition of the \Cod\
ions is systematically shifted towards higher temperature. This
strong pressure dependence is also reflected in pronounced
Schottky anomalies in the thermal-expansion coefficients, making
this quantity a sensitive probe to analyze spin-state transitions.
The microscopic origin of the pressure dependence is the volume
change due to the partial occupation of the $e_g$ levels of the
\Cod\ ions in the excited spin state in contrast to the empty
$e_g$ levels of LS \Cod\ ions. Another consequence of this volume
difference is a strong suppression of the phonon heat transport,
which sets in above the onset temperature of the spin-state
transition, because the random occupation of the Co sites with
\Cod\ ions in different spin states, i.\,e.\ of different sizes,
causes additional lattice disorder. By considering this disorder
as point-like defects, our analysis clearly shows this effect in
the temperature dependence of the corresponding scattering rates
of \lco, \pco\ and \nco, where the spin-state transition sets in
below room temperature, in contrast to \eco\ which exhibits a
stable LS state up to about 400\,K.

Quantitatively, the temperature dependence of the additional
scattering due to the spin-state transition can be described in a
model based on the Nordheim rule, which is usually applied to
describe scattering in disordered mixed alloys. For the
cobaltates, this means that the temperature dependence of this
additional scattering rate is given by the product of the
occupation numbers of the LS and of the excited spin states.
Moreover, we find that the scattering strength is constant for
\lco, \pco\ and \nco. Our analysis does, however, not allow to
distinguish between the different models discussed for the
spin-state transition in \lco, namely a LS\,--\,IS scenario with a
temperature-independent energy gap or the more recently proposed
model based on a spin-orbit coupled high-spin state with a
temperature-dependent energy gap.

The above described model explains a damping of the phonon heat
transport above the onset temperature of the spin-state
transition. Experimentally, however, we find that as a function of
increasing size of the $R^{3+}$ ions the thermal conductivity
decreases also in the temperature range below the respective onset
of the spin-state transition. By studying five different \lco\
single crystals, we could show that the low-temperature thermal
conductivity inversely scales with the amount of magnetic
impurities. Most likely, these impurities arise from a weak oxygen
nonstoichiometry giving rise to a certain amount of Co$^{2+}$ or
Co$^{4+}$ ions. Since it is known that these magnetic Co ions tend
to form so-called high-spin polarons by inducing a spin-state
transition in the neighbouring Co$^{3+}$ ions, our data suggest
that the phonon damping at low temperature arises from such a
doping-induced spin-state transition. The drastic increase of the
energy gap to the excited spin state of \Cod\ for decreasing
$R^{3+}$ ionic radii makes the formation of spin polarons more and
more unlikely, and thus can naturally explain the systematic
increase of the thermal conductivity in the temperature range
below the temperature-induced spin-state transition.

Additional low-temperature features have to be considered in \pco,
because the thermal expansion exhibits another Schottky anomaly
and there is an additional suppression of the thermal
conductivity. Both effects are not related to the
temperature-dependent spin-state transition of \Cod, but stem from
the crystal-field splitting of the ground-state multiplet of the
$4f$ shell of Pr$^{3+}$. Our analysis demonstrates that the
Schottky anomaly of the thermal expansion arises from a thermal
occupation of the first excited crystal-field level, whereas the
phonon damping is a consequence of resonant phonon scattering
between the first two excited crystal-field levels.

\section*{Acknowledgements}
We acknowledge useful discussions with M.~Haverkort and
A.~Sologubenko. This work was supported by the Deutsche
Forschungsgemeinschaft through Sonderforschungsbereich~608.

\section*{References}

\begin{thebibliography}{42}

\bibitem{jonker53a}
Jonker G H and Van Santen J H 1953 {\it Physica} {\bf XIX} 120

\bibitem{goodenough65a}
Goodenough J B and Raccah P M 1965 {\it J.\ Appl.\ Phys.} {\bf 36} 1031

\bibitem{senaris95a}
Se{\~n}ar\'{\i}s-Rodr\'{\i}guez M A and Goodenough J B 1995 {\it J.\ of Solid State Chem.} {\bf 116} 224

\bibitem{saitoh97b}
Saitoh T, Mizokawa T, Fujimori A, Abbate M, Takeda Y and Takano M 1997 {\it Phys.\ Rev.} B {\bf 55} 4257

\bibitem{asai98a}
Asai K, Yoneda A, Yokokura O, Tranquada J M, Shirane G and Kohn K 1998 {\it J.\ Phys.\ Soc.\ Japan} {\bf 67} 290

\bibitem{tokura98a}
Tokura Y, Okimoto Y, Yamaguchi S, Taniguchi H, Kimura T and Takagi H 1998 {\it Phys.\ Rev.} B {\bf 58} 1699

\bibitem{yamaguchi97a}
Yamaguchi S, Okimoto Y and Tokura Y 1997 {\it Phys.\ Rev.} B {\bf 55} 8666

\bibitem{kobayashi00b}
Kobayashi Y, Fujiwara N, Murata S, Asai K and Yasuoka H 2000 {\it Phys.\ Rev.} B {\bf 62} 410

\bibitem{sato08a}
Sato K, Bartashevich M I, Goto T, Kobayashi Y, Suzuki M, Asai K, Matsuo A and Kindo K 2008 {\it J.\ Phys.\ Soc.\ Japan} {\bf 77} 024601

\bibitem{zobel02a}
Zobel C, Kriener M, Bruns D, Baier J, Gr\"uninger M, Lorenz T, Reutler P and Revcolevschi A 2002 {\it Phys.\ Rev.} B {\bf 66} 020402

\bibitem{baier05a}
Baier J, Jodlauk S, Kriener M, Reichl A, Zobel C, Kierspel H, Freimuth A and Lorenz T 2005 {\it Phys.\ Rev.} B {\bf 71} 014443

\bibitem{noguchi02a}
Noguchi S, Kawamata S, Okuda K, Nojiri H and Motokawa M 2002 {\it Phys.\ Rev.} B {\bf 66} 94404

\bibitem{haverkort06a}
Haverkort M W, Hu Z, Cezar J C, Burnus T, Hartmann H, Reuther M, Zobel C, Lorenz T, Tanaka A, Brookes N B, Hsieh H H, Lin H J, Chen C T and Tjeng L H 2006 {\it Phys.\ Rev.\ Lett.} {\bf 97} 176405

\bibitem{podlesnyak06a}
Podlesnyak A, Streule S, Mesot J, Medarde M, Pomjakushina E, Conder K, Tanaka A, Haverkort M W and Khomskii D I 2006 {\it Phys.\ Rev.\ Lett.} {\bf 97} 247208

\bibitem{kriener04a}
Kriener M, Zobel C, Reichl A, Baier J, Cwik M, Berggold K,
Kierspel H, Zabara O, Freimuth A and Lorenz T 2004 {\it Phys.\
Rev.} B {\bf 69} 094417

\bibitem{kriener08a}
Kriener M, Braden M, Kierspel H, Senff D, Zabara O, Zobel C and Lorenz T 2008 {\it Preprint arXiv:0801.4188v1 [cond-mat.str-el]}

\bibitem{yan04a}
Yan J Q, Zhou J S and Goodenough J B 2004 {\it Phys.\ Rev.} B {\bf 69} 134409

\bibitem{berggold05a}
Berggold K, Kriener M, Zobel C, Reichl A, Reuther M, M\"{u}ller R,
Freimuth A and Lorenz T 2005 {\it Phys.\ Rev.} B {\bf 72}, 155116


\bibitem{berggold06a}
Berggold K, Lorenz T, Baier J, Kriener M, Senff S, Roth H, Severing A, Hartmann H, Freimuth A, Barilo S and Nakamura F 2006 {\it Phys.\ Rev.} B {\bf 73} 104430

\bibitem{pott83a}
Pott R and Schefzyk R 1983 {\it J.\ Phys.\ E -- Sci.\ Instrum.} {\bf 16} 444

\bibitem{stolen97a}
St{\o}len S, Gr{\o}nvold F, Brinks H, Atake T and Mori H 1997 {\it Phys.\ Rev.} B {\bf 55} 14103

\bibitem{yamaguchi96b}
Yamaguchi S, Okimoto Y and Tokura Y 1996 {\it Phys.\ Rev.} B {\bf 54} R11022

\bibitem{lorenz07a}
Lorenz T, Stark S, Heyer O, Hollmann N, Vasiliev A, Oosawa A and Tanaka H 2007 {\it J.\ Magn.\ Magn.\ Mat.} {\bf 316} 291

\bibitem{lorenz08a}
Lorenz T, Heyer O, Garst M, Anfuso F, Rosch A, R\"uegg Ch and Kr\"amer K 2008 {\it Phys.\ Rev.\ Lett.} {\bf 100} 067208

\bibitem{asai97a}
Asai K, Yokokura O, Suzuki M, Naka T, Matsumoto T, Takahashi H, M\^ori N and Kohn K 1997 {\it J.\ Phys.\ Soc.\ Japan} {\bf 66} 967

\bibitem{johannsen05a}
Johannsen N, Vasiliev A, Oosawa A, Tanaka H and Lorenz T 2005 {\it Phys.\ Rev.\ Lett.} {\bf 95} 017205

\bibitem{anfuso08a}
Anfuso F, Garst M, Rosch A, Heyer O, Lorenz T, R\"uegg Ch and
Kr\"amer K 2008 {\it Preprint arXiv:0803.1072v1 [cond-mat.str-el]}

\bibitem{yamaguchi96a}
Yamaguchi S, Okimoto Y, Taniguchi H and Tokura Y 1996 {\it Phys.\ Rev.} B {\bf 53} 2926

\bibitem{maignan04b}
Maignan A, Caignaert V, Raveau B, Khomskii D and Sawatzky G 2004 {\it Phys.\ Rev.\ Lett.} {\bf 93} 026401

\bibitem{lengsdorf04a}
Lengsdorf R, Ait-Tahar M, Saxena S S, Ellerby M, Khomskii D I,
Micklitz H, Lorenz T and Abd-Elmeguid M M 2004 {\it Phys.\ Rev.} B {\bf 69} 140403(R)

\bibitem{bermann76a}
Bermann R 1976 {\it Thermal Conduction in Solids} Clarendon Press Oxford

\bibitem{li07a}
Li S Y, Bonnemaison J B, Payeur A, Fournier P, Wang C H, Chen X H and Taillefer L 2007 {\it Preprint arXiv:0709.3075v1 [cond-mat.supr-con]}

\bibitem{kordonis06a}
Kordonis K, Sologubenko A V, Lorenz T, Cheong S W and Freimuth A 2006 {\it Phys.\ Rev.\ Lett.} {\bf 97} 115901

\bibitem{murata99a}
Murata S, Isida S, Suzuki M, Kobayashi Y, Asai K and Kohn K 1999 {\it Physica} B {\bf 263-264} 647--249

\bibitem{nordheim31a}
Nordheim L 1931 {\it Annalen der Physik} {\bf 9} 607

\bibitem{feldmann75a}
Feldmann K, Hennig K, Kaun L, Lippold B, Lukina M M, Matthies S, Matz W and Warming E 1975 {\it phys.\ stat.\ sol.} (b) {\bf 72} 817

\bibitem{podlesnyak94a}
Podlesnyak A, Rosenkranz S, Fauth F, Marti W, Scheel H J and Furrer A 1994 {\it J.\ Phys. -- Condens.\ Matter} {\bf 6} 4099

\bibitem{rosenkranz99a}
Rosenkranz S, Medarde M, Fauth F, Mesot J, Zolliker M, Furrer A, Staub U, Lacorre P, Osborn R, EcclestonR S and Trounov V  1999 {\it Phys.\ Rev.} B {\bf 60} 14857

\bibitem{slack71a}
Slack G A and Oliver D W 1971 {\it Phys.\ Rev.} B {\bf 4} 592

\bibitem{hofmann01a}
Hofmann M, Lorenz T, Uhrig G S, Kierspel H, Zabara O, Freimuth A,
Kageyama H, Ueda Y 2001 {\it Phys.\ Rev.\ Lett.} {\bf 87} 047202

\bibitem{berggold07a}
Berggold K, Baier J, Meier D, Mydosh J A, Lorenz T, Hemberger J, Balbashov A, Aliouane N and Argyriou D N 2007 {\it Phys.\ Rev.} B {\bf 76} 094418

\bibitem{orbach61a}
Orbach R 1961 {\it Proc.\ Roy.\ Soc.\ Lond.} {\bf 77} 821

\end{thebibliography}

\end{document}